\documentclass[conference]{IEEEtran}
	\IEEEoverridecommandlockouts
     \usepackage{fancyhdr,lipsum}

    \fancypagestyle{mahmood}{%
   
   \fancyhead[C]{As seen in 8th IEEE/IFIP International Workshop On Analytics For Network And Service Management, May 2023, Miami, Florida, USA.}
    }%

    \makeatletter
    \let\ps@IEEEtitlepagestyle\ps@mahmood
    \makeatother
	\usepackage{cite}
	\usepackage{amsmath,amssymb,amsfonts}
	\usepackage[noend]{algorithmic}
	\usepackage[dvisvgm, usenames, dvipsnames]{color}
	\usepackage{graphicx}
	\usepackage{textcomp}
	\usepackage{xcolor}
	\usepackage{verbatim}
	\usepackage{afterpage}
	\usepackage{ragged2e}
	\usepackage{comment}

    \def\BibTeX{{\rm B\kern-.05em{\sc i\kern-.025em b}\kern-.08em T\kern-.1667em\lower.7ex\hbox{E}\kern-.125emX}}
    
	\usepackage[margin=1in,footskip=0.25in]{geometry}
\begin{document}
	
		\title{Data-Centric Machine Learning Approach for Early Ransomware Detection and Attribution}
	\author{\IEEEauthorblockN{A. Vehabovic$^{1}$, H. Zanddizari$^{2}$, N. Ghani${^1}$, F. Shaikh${^1}$, E. Bou-Harb$^{2}$, M. Safaei Pour${^3}$, J. Crichigno$^{4}$ \\
	\textit{$^{1}$Univ. of South Florida, $^{2}$Univ. of Texas San Antonio, $^{3}$San Diego State Univ., $^{4}$Univ. of South Carolina}}
	}
	\maketitle

\maketitle

\begin{abstract}
Researchers have proposed a wide range of ransomware detection and analysis schemes. However, most of these efforts have focused on older families targeting Windows 7/8 systems. Hence there is a critical need to develop efficient solutions to tackle the latest threats, many of which may have relatively fewer samples to analyze. This paper presents a \textit{machine learning} (ML) framework for early ransomware detection and attribution. The solution pursues a data-centric approach which uses a minimalist ransomware dataset and implements static analysis using \textit{portable executable} (PE) files. Results for several ML classifiers confirm strong performance in terms of accuracy and zero-day threat detection.
\end{abstract}
%

\section{Introduction}
Ransomware operates by encrypting files on a host computer and demanding some form of payment to release the keys. This malware has become the most lucrative revenue source for cybercriminals, and many ransomware ``families'' have impacted a wide range of users.  Moreover, numerous cyber-criminal affiliates are also offering \textit{ransomware-as-a-service} (RaaS), further reducing the barrier to such extortion \cite{mouss2021}.

Ransomware follows a multi-stage ``kill-chain'' comprising of reconnaissance, distribution, installation, communication, encryption, and extortion \cite{vehabovic2022},\cite{berrueta2019}. To date, numerous designs have been evolved with increasing levels of secrecy, speed, and complexity. For example, various methods have been used to breach systems (e.g., remote access, drive-by, and privilege escalation) and encrypt data in collaboration with \textit{command and control} (C\&C) servers. Data exfiltration has also been used to extort users (double ransomware) \cite{vehabovic2022}. As this threat continues to grow, surveys indicate that almost half of large corporations have experienced such attacks \cite{kapoor2022}. Windows ransomware is of particular concern as this \textit{operating system} (OS) is still the most prevalent.

In light of the above, researchers have proposed a range of ransomware analysis solutions. Many of these schemes extract information from network traces or host files/logs to train advanced \textit{machine learning} (ML) classifiers. However, most efforts have focused on a specific ransomware family or older families targeting dated Windows 7/8 systems. As such, these methods may not be applicable to the latest threats facing Windows 10/11 users. Hence there is a pressing need to detect new ransomware designs \textit{and} classify them for improved mitigation, i.e., attribution. Preferably, ransomware should be tackled early in the kill-chain to minimize damage \cite{mouss2021}. Since new ransomware releases will likely have fewer available samples, solutions must also operate effectively with smaller ``minimalist'' datasets. This requirement is very much in line with current trends in \textit{artificial intelligence} (AI) to develop more focused ``data-centric'' solutions \cite{ng2022}.

Accordingly, this paper presents a novel ML solution for ransomware detection and attribution using static analysis. First, a unique malware repository is built by collecting samples of some of the latest ransomware families, i.e., Babuk/Babyk, BlackCat, Chaos, DJVu/STOP, Hive, LockBit, Netwalker, Sodinokibi/REvil, and WannaCry (after 2017). Next, feature extraction is done using Windows \textit{portable executable} (PE) format file information. Finally, several supervised ML classifiers are trained and tested on these extracted features, including \textit{support vector machines} (SVM), \textit{random forest} (RF), \textit{extreme gradient boost} (XGBoost), and \textit{feed-forward neural networks} (FNN) \cite{geron2019}. Overall, this solution has very amenable run-times and can be integrated into network/host-based defenses to target ransomware early in the kill-chain (prevention).

This paper is organized as follows. Section \ref{survey} reviews some key studies on ransomware analysis. Next, Section \ref{ML_solution} details the proposed ML-based framework, including dataset collection and feature extraction. Performance results are then presented in Section \ref{performance}, followed by future work directions in Section \ref{conclusions}.

\section{Literature Review}
\label{survey}
A range of ransomware analysis schemes have been proposed, and survey articles have detailed various (overlapping) taxonomies to classify these methods, e.g., static or dynamic analysis, network- or host-based, etc \cite{mouss2021}-\cite{berrueta2019}. These efforts are further reviewed here.

Static analysis examines executable files to detect artifacts of maliciousness, e.g., via author attribution, code/segment identification (de-anonymization), etc \cite{mouss2021}. Some common methods used here include \textit{binary code analysis} (BCA), source code analysis via reverse engineering, and C\&C server domain prediction \cite{vehabovic2022}. For example, \cite{poudyal2018} specifies a multi-level framework to detect ransomware from raw binaries, assembly code, and libraries. ML classifiers are then trained with the extracted data, yielding detection rates around 90\%.  Meanwhile, \cite{zhang2019} transforms code sequences into N-grams and extracts frequency-based features for classification. Results show detection rates around 91\% for several ML classifiers (decision tree, RF, etc). However, code-based analysis is very labor-intensive \cite{Mulders2017} and represents a more latent ``post-infection'' forensics approach.

Recent efforts have also used other static features to analyze ransomware. For example, \cite{wang2021} leverages image processing techniques to convert ransomware binary files into grayscale images and then performs texture analysis for feature extraction. Results for several ML classifiers show high accuracy (97\%) for a small dataset with a mix of old and new ransomwares (379 samples). However, this scheme imposes added computational burdens and does not consider benign applications. Meanwhile, \cite{zhu2022} details another static analysis scheme which extracts entropy and image-based features to train a specialized Siamese NN classifier. Tests with a small dataset (about 1,000 samples and 10 families) show accuracy values in the mid-90\% range but notably lower precision and recall rates (upper 70\% range). Also, most of the ransomware families used here are older (mid-2010s) and benign applications are not considered.

Studies have also used static PE header file data for broader malware detection (not just ransomware). However, these efforts focus on detection and not attribution.  For example, \cite{kim2016} collects many samples (over 100,000) from a repository called {\tt VX Heaven} (now inactive) and trains ML classifiers using 7-10 extracted PE file features. Results show detection rates in the upper 90\% range. Also, \cite{liao2021} extracts PE features from about 5,500 malware samples and 1,200 benign applications (early 2010s). Detection is done using a set of heuristics, achieving 95\% accuracy. Finally, \cite{rezaei2020} extracts 9 PE file features (on sections, data directories, and entropy) from a dataset with 1,200 malicious and benign samples each. Results for several classifiers show 95\% detection rates. However, these studies present no details on their malware datasets, most of which are over a decade old.

By contrast, dynamic analysis scans run-time actions and event sequences for ransomware activity. Specifically, dynamic network-based schemes examine packet traces for C\&C communications, \textit{domain name service} (DNS) queries, network storage access, etc. For example, \cite{almash2019} presents a detection system for Locky ransomware which uses traffic features to train classifiers and yields over 95\% detection rates.  Meanwhile, \cite{morato2018} analyzes \textit{server message block} (SMB) protocol patterns to detect older ransomware (2015-2017). The NetConverse scheme \cite{alhawi2018} also uses ML methods to analyze host traffic for earlier threats and achieves high detection rates (over 95\%). Finally, \cite{roy2021} uses deep learning to analyze network activity and classify abnormal operation in Windows 7. Results show high detection rates for several families (over 97\%).

Meanwhile, dynamic host-based schemes monitor local system activity to detect ransomware, e.g., memory and file operations, \textit{application programmer interface} (API) function calls, \textit{dynamic link library} (DLL) calls, etc. For example, \cite{kharraz2016} uses a sandbox to track file encryption/deletion, persistent messages, etc. Results show 96\% detection rates for older ransomware types (mid-2010s). Also, \cite{kolodenker2017} presents a scheme to monitor and store encryption keys for ransomware detection and file recovery. Results show successful mitigation of 12 out of 20 families. 
Similarly, \cite{kharraz2017} scans input/output requests for ransomware activity and flags affected files. Studies have also proposed ransomware ``paranoia'' schemes that try to detect environments and avoid fingerprinting/detection, e.g.,\cite{alsabeh2020} tracks API calls.

Although the above works present some notable contributions, key concerns still remain. Foremost, studies have largely focused on older ransomware targeting Windows 7/8 systems (mid-2010s). Given the expanding nature of this threat, it is imperative to study newer families targeting Windows 10/11. However, there are few datasets here, and new malwares may have smaller sample sizes to analyze (a challenge for ML schemes). Hence effective ``data-centric'' \cite{ng2022} schemes are required for minimalist datasets. Finally, ransomware detection and attribution schemes must have amenable run-times and preferably target ransomware earlier the distribution/delivery stages to minimize damage \cite{vehabovic2022}. It is here that static analysis offers an expedient approach for tackling malicious payloads prior to infection. By contrast, dynamic analysis requires more indepth examination of network or host activities over longer intervals in virtual environments. As a result, a static analysis solution is presented using PE format file analysis.
\begin{figure*}[h]
    \centering
    \includegraphics[width=6.25in, height=2.45in]{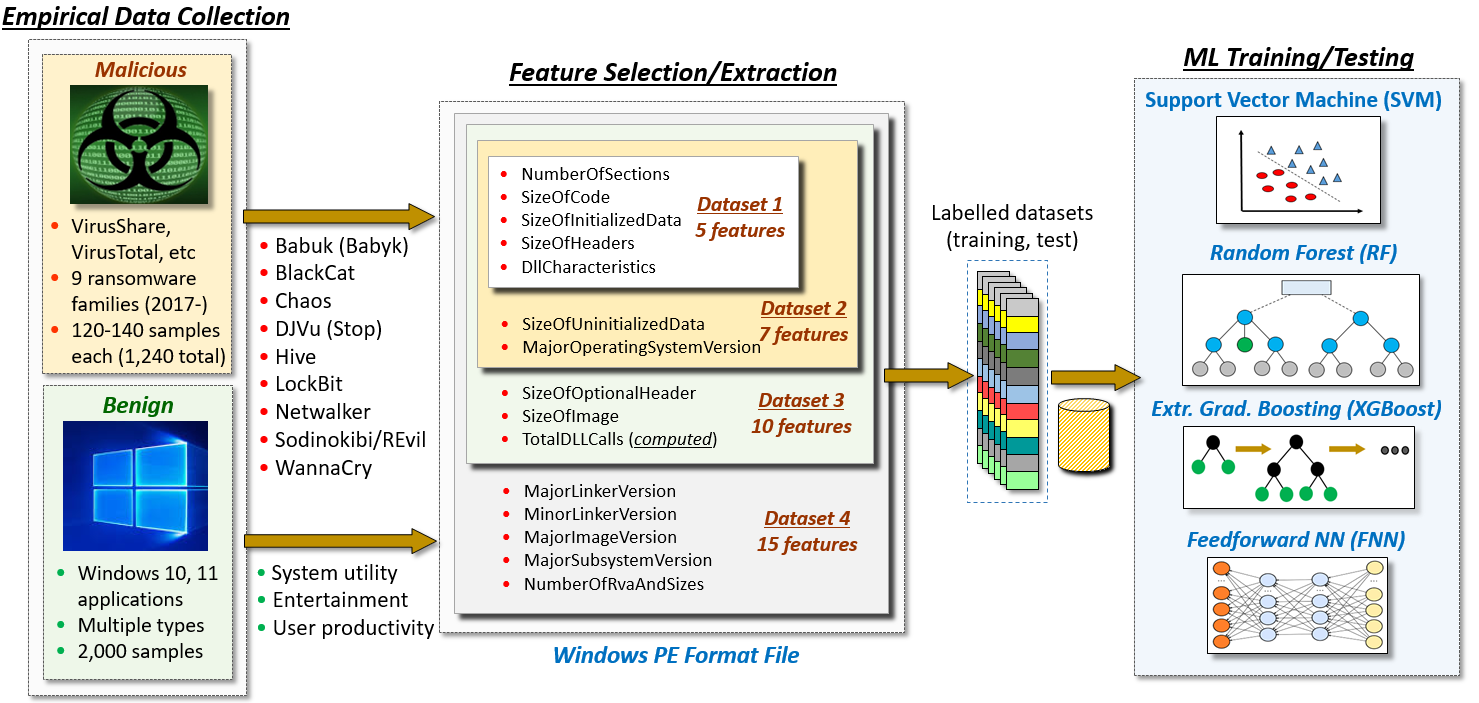}
    \caption{Overview of static analysis ML framework for ransomware detection and attribution}
    \label{ML_overview}
\end{figure*}
\section{Data-Centric Static Analysis using ML}
\label{ML_solution}
The static analysis framework for ransomware detection and attribution (classification) is shown in Fig. \ref{ML_overview} and comprises of several stages. The first stage (Empirical Data Collection) builds an up-to-date repository of some of the latest Windows 10/11 ransomware threats (since 2017). Regular benign Windows-based applications are also added here to improve classifier performance. The second stage (Feature Selection/Extraction) processes raw executables to extract key features. An efficient static analysis approach is proposed here using Windows PE format files. Finally, the last stage (ML Training/Testing) uses the feature datasets to train ML classifiers to detect \text{and} attribute ransomware. On a high level, this setup follows a well-defined ML approach, similar to that used in other studies. However, the novel contributions here include the collection of new ransomware datasets and extraction of lightweight static feature sets. Further details are now presented.
\begin{table}[h]
	\centering
\caption{Empirical dataset}
\label{table_dataset}
\begin{tabular}{|l|c|c|c|} 
\hline
\textbf{ Family } & \textbf{Samples} & \textbf{Avg. Size}  & \textbf{Avg. PE File} \\ \hline \hline
Babuk (Babyk) & 140 & 0.19 MB & 32.68 KB \\ \hline
BlackCat & 120 & 3.91 MB & 1,147 KB \\ \hline
Chaos & 140 & 0.49 MB & 35.2 KB \\ \hline
DJVu (STOP) & 140 & 0.71 MB & 66.2 KB \\ \hline
Hive & 140 & 3.51 MB & 403.9 KB \\ \hline
LockBit & 140 & 1.30 MB & 171.5 KB \\ \hline
Netwalker & 140 & 0.26 MB & 35.72 KB \\ \hline
Sodinokibi & 140 & 0.30 MB & 50.89 KB \\ \hline
WannaCry & 140 & 7.62 MB & 21.83 KB \\ \hline
\hline
Benign & 2,000 & 26.86 MB & 155.88 KB \\ \hline
\end{tabular}  
\end{table}

\subsection{Empirical Data Collection}
\label{raw_data}
As per Section \ref{survey}, existing studies on PE file analysis provide little/no details on their datasets, e.g., type of malwares, executable file sizes, collection time frames, percentage of ransomware, etc. Many of these malwares are old and related repositories are inactive \cite{kim2016}. Hence a new repository is curated for the latest ransomware families. Now given the rapidly changing nature of the ransomware threat, it may be difficult to get sufficient samples of each. Hence realistic ``data-centric'' ML frameworks must achieve good detection and attribution with minimalist datasets (perhaps only hundreds of samples). However, limited dataset size/diversity can also have a negative impact on classifier performance.

Now many active repositories host malware executables, e.g., {\tt MalwareBazar}, {\tt Triage}, {\tt VirusShare}, and {\tt VirusTotal}, etc. These sites provide varying degrees of access and usability, e.g., {\tt VirusTotal} and {\tt VirusShare} require registration to access uploads. Detailed cross-checking and analysis also shows notable duplication across portals, e.g., many Sodinokibi samples on {\tt MalwareBazar} match those on {\tt Triage}. There are also discrepancies between the number of samples for each family, e.g., DJVu is abundant whereas Babuk/Babyk and BlackCat are more scarce. Finally, some repositories ({\tt VirusShare} and {\tt VirusTotal}) do not organize or label their data, further complicating collection. Hence unlabeled data dumps have to be tediously analyzed using hashing and cross-checked with labelled samples. Hence there is potential for a lack of diversity, even scarcity, of new ransomware.

In light of the above, a smaller ``minimalist'' data repository is curated with 9 active ransomware families, i.e., Babuk/Babyk, BlackCat, Chaos, DJVu/STOP, Hive, LockBit, Netwalker, Sodinokibi/REvil, and WannaCry (Table \ref{table_dataset}).  These families are amongst the most prevalent ransomware threats in 2022, as per the IBM X-Force Threat Intelligence Index, i.e., LockBit (17\%) followed by WannaCry (11\%) and BlackCat (9\%). A total of 140 unique executables are collected for each family, except for BlackCat which only yielded 120 samples due to scarcity, i.e., total of 1,240 malicious samples. Many Windows 10/11 applications are also added to construct a benign class (2,000 samples). These programs are collected from a range of websites and include system utility, entertainment, and productivity tools (Fig. \ref{ML_overview}). Overall, having a large set of non-malicious training data is very beneficial since regular applications downloads will exceed (unintended) ransomware downloads, This addition contrasts with work in \cite{wang2021},\cite{zhu2022}.

\subsection{Feature Selection/Extraction}
\label{feature_extraction}
ML classifier performance is heavily dependent upon input training data. Hence feature extraction (engineering) plays a vital role in transforming raw executables to generate meaningful information for classifiers \cite{geron2019}. As per Section \ref{survey}, static analysis is more expedient for tracking ransomware early in its kill-chain.  Hence this strategy is applied to Windows PE format files which contain data structures to support program execution in 32-bit and 64-bit Windows OS environments. Namely, these files use the \textit{common object file format} (COFF) and contain information for the OS loader to setup/run wrapped executable code (including memory mapping and permissions). For example, a PE format file has several initial lead-in headers along with multiple sections. Here each section specifies file content (i.e., code or data) and also contains its own section header.

As per Section \ref{survey}, studies on PE format files have considered a range of malwares for Windows 7/8 \cite{kim2016}-\cite{rezaei2020} (mostly unspecified and not necessarily ransomware). Hence there is a further need to extend such analysis to Windows 10/11 ransomware threats. Now PE files contain a wealth of information, and programs can have unique non-overlapping parameters (depending upon functionality). Hence when extracting PE format data, it is important to select a subset of parameters which exist across all sample files and also exhibit good variability.

In light of the above, PE files are generated for all exectuables, with the resultant sizes shown in Table \ref{table_dataset}. A total of 4 datasets are built by extracting feature vectors with 5, 7, 10, and 15 parameters, labeled as Datasets 1-4, respectively (Fig. \ref{ML_overview}). Each successive vector expands upon its predecessor by adding new parameters. Now the exact parameters are chosen using careful experimentation with the \textit{Image\_File\_Header}, \textit{Image\_Optional\_Header}, and \textit{Image\_Section\_Header} sections. Some key features include \textit{NumberOfSections}, \textit{SizeOfCode}, \textit{SizeOfHeaders}, etc. Note that PE files also contain information on \textit{dynamic-link library} (DLL) calls which are indicative of functionality. For example, ransomware typically calls encryption, socket communications, and registry-modification functions. Hence the total number of DLL calls is also added to the 10 and 15 feature vectors (\textit{TotalDLLCalls}, Fig. \ref{ML_overview}). Note that this is a \textit{computed} feature and not an extracted parameter.

\section{Performance Evaluation}
\label{performance}
The  static analysis framework is now evaluated using the data repository from Section \ref{raw_data}. Namely, feature vectors extracted from the PE files are labelled to generate input datasets. These datasets are then used to train/test supervised ML classifiers, i.e., SVM, RF, XGBoost, and FNN \cite{geron2019} (Fig. \ref{ML_overview}). All evaluation is done using the {\tt Keras} and {\tt TensorFlow} toolkits, as well as {\tt Pandas} and {\tt Sklearn}. As per Section \ref{raw_data}, a total of 9 malicious ransomware families are evaluated along with a set of benign applications, i.e., 10 classes. As noted earlier, there are a total of 1,240 malicious samples (140 samples for each family except BlackCat which has 120 samples).  The samples for each class are further partitioned to generate separate training and testing pools. Namely, 20 random samples of each class are selected for testing and the remainder are used for training, i.e., 120 training samples for all classes except BlackCat which only has 100 samples.  Furthermore, 1,700 benign samples are selected for training and the remaining 300 samples are used for testing. This partitioning reflects an approximate 85/15 training/testing split. All results are averaged over 100 trial runs, with each using a different randomized 85/15 partitioning of the datasets. Detailed findings are now presented.
\begin{figure}[h]
	\centering
	\includegraphics[width=3.05in, height=1.75in]{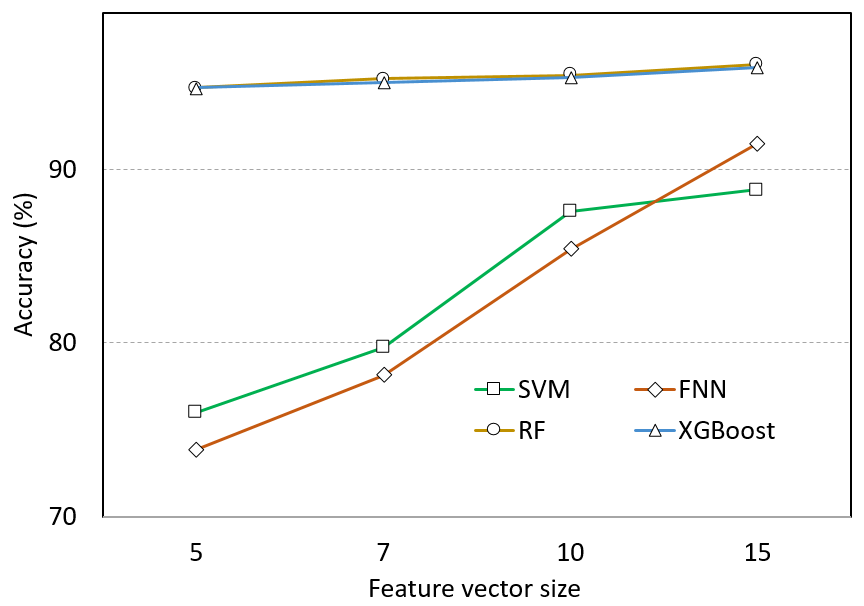}
	\caption{Average multi-class accuracy (100 trials)} 
	\label{fig_avg_accuracy}
\end{figure}

The average accuracy values (over all runs) are plotted for different feature vector sizes in Fig. \ref{fig_avg_accuracy}, i.e., multi-class attribution. Results show improved performance for all schemes with increasing feature vector sizes. In particular, the SVM and FNN classifiers give the best improvement, with accuracy gains of 15-20\%.  Conversely, the RF and XGBoost classifiers have much lower gains as feature vector sizes increase from 5 to 15 parameters, i.e., 0.5-1.5\% range. These two classifiers also give the best accuracy (94-96\% range). However, the FNN scheme approaches these methods with 15 features, i.e., 91\% accuracy. These findings are very encouraging given the relatively small-sized training datasets and feature vectors used. The results also match those for other schemes using much heavier feature extraction and ML algorithms, e.g., image and entropy-based features, deep NN algorithms, etc \cite{wang2021},\cite{zhu2022}.

Next, consider attribution errors in more detail. Indeed, mis-classifying ransomware as benign is much more harmful than mis-classifying it as the wrong type of ransomware, i.e., since such errors can allow malware to bypass network or host defenses and infect host machines. Hence to quantify this behavior, a modified \textit{ransomware detection rate} (RDR) is defined as:
\begin{equation}
    RDR = \frac{T_{rs}}{T_{rs}+F_{rs}}
    \label{eq_RDR}
\end{equation}
where $T_{rs}$ is the total number of ransomware samples classified as (any class of) ransomware, and $F_{rs}$ is the total number of ransomware samples mis-classified as benign, i.e., total number of ransomware test samples is $(T_{rs}+F_{rs})$.  This metric essentially captures the \textit{binary} detection capability of a multi-class classifier and is similar to the recall formula, i.e., tracks false negatives. A \textit{benign detection rate} (BDR) is also defined as:
\begin{equation}
     BDR = \frac{T_{bn}}{T_{bn}+F_{bn}}
    \label{eq_BDR}
\end{equation}
where $T_{bn}$ is the total number of benign samples classified as benign, and $F_{bn}$ is the total number of benign samples mis-classified as ransomware. In general, though, false negative attribution of benign executables is less of a security concern.  
\begin{figure}[h]
	\centering
	\includegraphics[width=3.05in, height=1.75in]{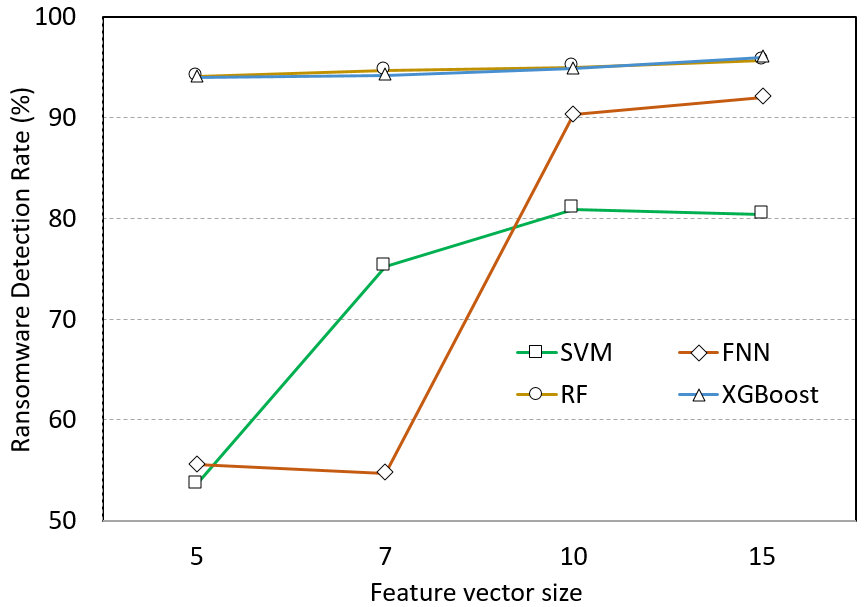}
	\caption{Average ransomware detection rate (100 trials)} 
	\label{fig_RDR}
\end{figure}
\begin{figure}[h]
	\centering
	\includegraphics[width=3.05in, height=1.75in]{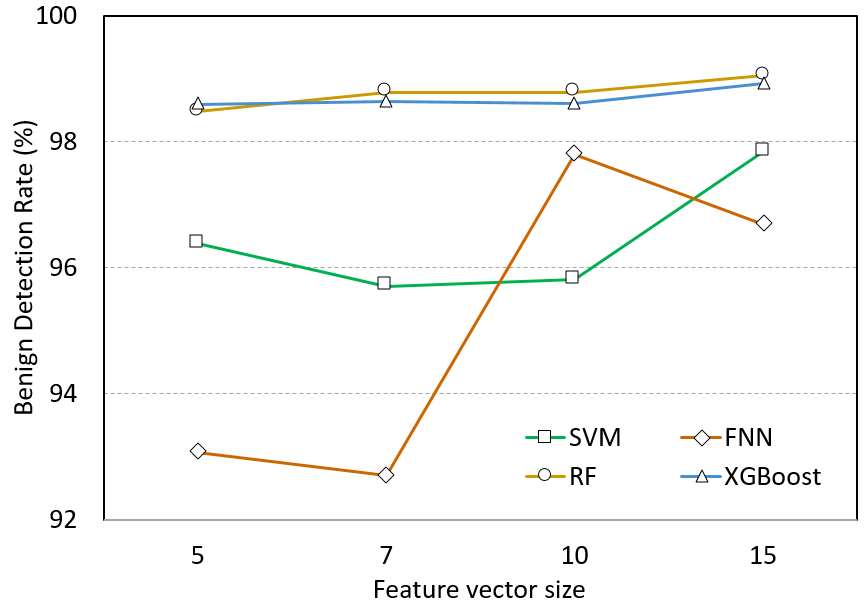}
	\caption{Average benign detection rate (100 trials)} 
	\label{fig_BDR}
\end{figure}

Accordingly, Fig. \ref{fig_RDR} plots the binary RDR results averaged over 100 trails. Akin to the multi-class case, the RF and XGBoost schemes give the highest ransomware detection rates, close to 95\%. Again, larger feature vector sizes also give smaller improvements with these classifiers, i.e., 2\% range. By contrast, the SVM and FNN schemes give very poor results for small feature vectors, with ransomware mis-classification rates around 50\% (1-$RDR$). These classifiers are also very sensitive to feature vector size. Nevertheless, the FNN scheme still approaches the performance of the RF and XGBoost schemes with larger feature vectors, i.e., 92\% RDR. The BDR results are also plotted in Fig. \ref{fig_BDR}. As expected, these values are higher than the RDR values since a larger amount of benign data is used for training. Again, the RF and XGBoost schemes give the lowest benign program mis-classification rates, close to 99\%.  Although the other methods (SVM, RF) give slightly lower BDR rates, they are still over 92\% (less than 1 error in 12). Note that these binary detection rates closely match those from other malware detection studies which make use of much larger datasets and more elaborate feature extraction schemes (Section \ref{survey}).

\begin{figure}[h]
	\centering
	\includegraphics[width=3.05in, height=2.05in]{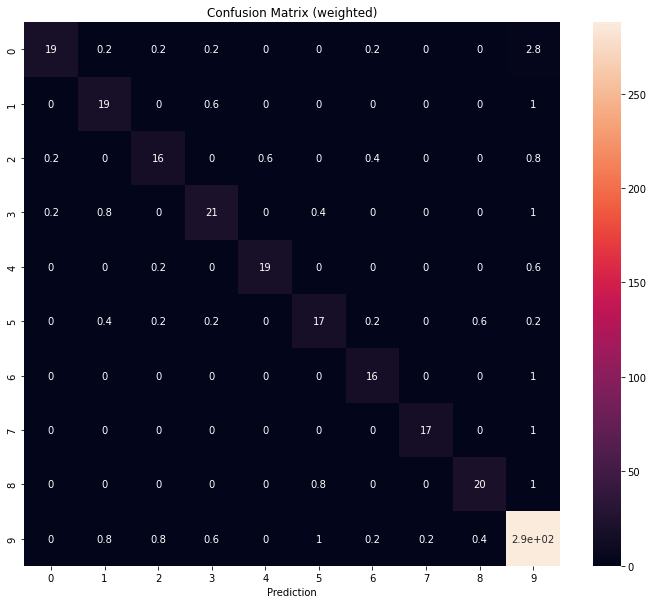}
	\caption{Confusion matrix (XGBoost, 15 features)} 
	\label{fig_confusion_matrix}
\end{figure}

Meanwhile, Fig. \ref{fig_confusion_matrix} shows an average confusion matrix for the XGBoost classifier (classes 0-8 represent the 9 ransomware families and class 9 represents the benign class). Here, the numbers in row 9 are larger as there are more benign test samples. These results confirm that most samples are classified correctly, i.e., diagonal numbers dominate.  Moreover, even when ransomware samples are mis-classified, they are mostly flagged as another ransomware (mirroring RDR results in Fig. \ref{fig_RDR}).
\vspace{-0.1in}

Tests are also done to gauge zero-day attack detection. Namely, 8 out of the 9 ransomware families are aggregated into a single malicious class and used to train a binary classifier (versus benign class). The remaining family is then tested as a zero-day threat, i.e., to see if the binary classifier can flag it as ransomware. Hence all samples are either used for training or testing. The associated detection rates are shown in Table \ref{table_zero_day} for all possible zero-day attack scenarios.  These results show very good performances for several classifiers. For example, the RF (XGBoost) scheme gives approximately 80-99\% detection rates for 8 (6) out of the 9 ransomwares tested, i.e., at least 4 out of 5 zero-day attacks detected. However, the WannaCry malware is very effective at evading all schemes and has low detection rates in 14-20\% range, i.e., only 1 in 7 detected. Hence additional PE file features (parameters) or other static parameters may need to be incorporated.

\begin{table}[h]
	\centering
\caption{Zero-day attacks (detection accuracy)}
\label{table_zero_day}
\begin{tabular}{|l|c|c|c|c|} 
\hline
\textbf{ Zero-Day } & \textbf{ SVM} & \textbf{ RF }  & \textbf{XGB} & \textbf{ FNN } \\ \hline \hline
Babuk/Babyk & 81.43\% & 94.86\% & 93.57\% & 74.38\% \\ \hline
BlackCat & 36.67\% & 78.17\% & 35.83\% & 65.70\% \\ \hline
Chaos & 93.57\% & 87.00\% & 87.14\% & 83.95\% \\ \hline
DJVu/STOP & 82.86\% & 98.64\% & 90.71\% & 79.40\% \\ \hline
Hive & 15.71\% & 82.71\% & 82.88\% & 72.30\% \\ \hline
LockBit & 57.14\% & 84.79\% & 62.86\% & 63.06\% \\ \hline
NetWalker & 95.00\% & 97.36\% & 97.14\% & 98.21\% \\ \hline
Sodinokibi/REvil & 95.00\% & 91.00\% & 90.05\% & 89.86\% \\ \hline
WannaCry & 13.57\% & 14.86\% & 14.29\% & 19.81\% \\ \hline
\end{tabular}  
\end{table}

Finally, run-times are measured by averaging PE format file generation, feature extraction, and ML attribution times. Tests are done on a Windows 11 server with a 3.60 GHz {\tt Intel Core i9} processor and 64 Gb of \textit{random access memory} (RAM). Only \textit{trained} ML models are timed to reflect operational settings, and PE file generation is done using a {\tt Github} package (\textit{https://github.com/erocarrera/pefile}). Results show that PE file generation times are directly correlated to executable file sizes (Table \ref{table_dataset}). For example, benign files average 1.46 sec, whereas ransomware files range from 50-300 ms (larger for LockBit, DJVu/STOP, and Hive). Meanwhile, classification times vary between 2-8 ms for the SVM, RF, and XGboost classifiers, but are higher for the FNN scheme at 47 ms. Overall, many operators are willing to accept these delays for scanning incoming downloads/attachments in network/host-based defenses.

\section{Conclusions}
\label{conclusions}
It is imperative to track the latest ransomware releases and develop effective solutions for mitigating these threats. This paper presents a static analysis scheme for ransomware detection and attribution. First, a new dataset is curated with the latest Windows 10/11 ransomware families. Windows \textit{portable executable} (PE) format files are then used to extract feature vectors and train \textit{machine learning} (ML) classifiers. Overall findings show very good performance in terms of ransomware detection, attribution, and zero-day threat detection. These results are achieved using minimalist datasets with about 100-120 training samples per class and relatively compact feature vectors. The solution also gives very amenable run-times for realistic settings. This work presents a strong basis from which to expand into future work. Specifically, more ransomware families can be added to the repository and refined feature extraction and ML methods can also be studied. 

\bibliographystyle{IEEEtran}
\bibliography{references.bib}

\end{document}